\documentclass[journal]{IEEEtran}
\usepackage{mathrsfs}
\usepackage{amsmath,amsfonts}
\allowdisplaybreaks[4]
\usepackage{autobreak}
\usepackage{algorithmic}
\usepackage{algorithm}
\usepackage{xspace} 
\usepackage{amssymb}
\usepackage{array}
\usepackage[caption=false,font=normalsize,labelfont=sf,textfont=sf]{subfig}
\usepackage{textcomp}
\usepackage{stfloats}
\usepackage{enumitem}
\usepackage{url}
\usepackage{verbatim}
\usepackage{graphicx}
\usepackage{cite}
\usepackage{xcolor}
\definecolor{deepgreen}{rgb}{0, 0.6, 0.2}
\usepackage[hidelinks, colorlinks=true, linkcolor=blue, citecolor=deepgreen]{hyperref}
\begin{document}

\title{ Authenticated Sublinear Quantum Private Information Retrieval }

\author{Fengxia Liu, Zhiyong Zheng,  Oleksiy Zhedanov,  Yi Zhang, Heng Guo,    Kun Tian*,  Zixian Gong*, Zhiming Zheng
\thanks{The work of Zhiyong Zheng was supported in part by the Major Project of Henan Province  under Grant 225200810036.
 The work of Fengxia Liu was supported in part by the Great Bay University  under Grant H24120002, H24120003. (Corresponding author: Kun Tian and Zixian Gong )

Fengxia Liu  is with Great Bay University, Dongguan, 523830, China(e-mail: shunliliu@gbu.edu.cn)

 Zhiyong Zheng  and Oleksiy Zhedanov  are with Great Bay University, Dongguan, 523830, China,   Engineering Research Center of Ministry of Education for Financial Computing and Digital Engineering, Renmin University of China, Beijing, 100872, China(e-mail: zhengzy@ruc.edu.cn). 
 
 Zhiming Zheng is with Beihang University, Beijing, 100191, China
 
Kun Tian,  Yi Zhang, Heng Guo, Zhe Hu and Zixian Gong are  with Engineering Research Center of Ministry of Education for Financial Computing and Digital Engineering, Renmin University of China, Beijing, 100872, China; Great Bay University, Dongguan, 523830, China (e-mail: gzx@ruc.edu.cn) }}

\markboth{}%
{Shell \MakeLowercase{\textit{et al.}}: A Sample Article Using IEEEtran.cls for IEEE Journals}

\maketitle

\begin{abstract}

this paper present a new lower bound on Quantum Private Information Retrieval (QPIR) communication, derived via quantum relative entropy and mutual information together with Uhlmann’s lemma and quantum Pinsker inequalities, which tightens classical entropy‐based results. Leveraging this bound, we design symmetric QPIR schemes that achieve sublinear communication while preserving information-theoretic security. First, by combining trapdoor claw‐free functions with localized CHSH tests, we create the first authenticated QPIR that against a malicious quantum server while using sublinear communication. Second, we optimize quantum homomorphic encryption so that a single server can search an encrypted superposition with only $O(\sqrt{N})$ communication. Third, in the multi‐server setting, we employ simple quantum Fourier transforms and classical index‐subset partitioning so that each server performs only basic Boolean operations and single‐qubit measurements, enabling data recovery with $O(\log n)$ communication and minimal hardware overhead. The work also analyzes security trade-offs under quantum specious adversaries, providing theoretical guarantees for privacy and correctness.

\end{abstract}


\section{Introduction}\label{sec1}

Private Information Retrieval (PIR) addresses a fundamental cryptographic challenge: enabling users to retrieve specific entries from a database without revealing which entries were accessed. Classical PIR protocols face inherent trade-offs between communication efficiency, security assumptions, and server architecture requirements. Information-theoretically secure single-server classical PIR necessitates linear communication complexity \cite{bennychor}, while multi-server schemes reduce overhead through database replication at the cost of requiring non-colluding servers \cite{beimel}.  Quantum PIR (QPIR) enhances this model by employing quantum states in place of classical bits for communication, thereby offering a theoretically superior and physically secure approach to data privacy. This enhancement is based on the principles of quantum superposition, entanglement, and the non-clonability of quantum information, which collectively expand the boundaries of privacy protection \cite{kon21}. Even if up to $t$ servers conspire, they remain incapable of discerning the users' query intentions.

\subsection{Quantum Advantages in PIR}
Quantum Private Information Retrieval (QPIR) exploits quantum resources to achieve unprecedented privacy guarantees. Initial breakthroughs demonstrated that multi-server QPIR protocols leveraging pre-shared entanglement can attain capacities reaching $\min\{1, 2(n-t)/n\}$ for $t$-private scenarios \cite{song21}, significantly surpassing classical bounds of $1/(1 + t/(n-t))$. This advantage stems from quantum superposition enabling simultaneous query encoding and entanglement facilitating secure channel establishment. Subsequent work \cite{legall} introduced quantum state compression techniques, achieving $\mathcal{O}(\sqrt{n})$ communication complexity through superposition-based queries  an exponential improvement over classical linear scaling. However, these protocols face critical security limitations under approximate privacy models \cite{broadbentPIR} and assume semi-honest server behavior \cite{Brakerski18}.

The single-server scenario presents particularly stringent challenges. Nayak's bound \cite{Nayak} establishes that even approximate QPIR requires $\Omega(n)$ quantum bits of communication, aligning with Holevo's theorem constraints on quantum information density. This fundamental limit persists across various security models, including specious adversaries \cite{broadbentPIR}. Hybrid approaches combining lattice-based cryptography with quantum techniques have emerged as promising alternatives, with Learning-with-Errors (LWE) based protocols \cite{liu} \cite{liu2} enabling sublinear complexity under computational assumptions while maintaining post-quantum security \cite{zheng2}. In this paper, we also explore the key differences between quantum PIR and classical PIR, and demonstrate how quantum PIR achieves strictly better communication complexity compared to classical PIR.

\subsection{Technical Challenges and Limitations}
Despite remarkable progress, QPIR development faces the following principal challenges:

\begin{enumerate}[label=\textbullet]
    \item \textbf{Fundamental Security Trade-offs}
    \begin{enumerate}[label=\arabic*.]
        \item No-go results for ideal protocols. Even in the quantum setting, perfect concealment of user queries leaves databases vulnerable to attacks \cite{william}.
        \item Lower bounds on communication. QPIR protocols require linear communication $\mathcal{O}(n)$ under information-theoretic security against plausible adversaries (e.g., specious adversaries) \cite{kon21}\cite{broadbentPIR}, negating claims of sublinear communication.
    \end{enumerate}

    \item \textbf{Restricted Adversary Models}
    \begin{enumerate}[label=\arabic*.]
        \item Honest server assumptions: Many protocols \cite{legall}\cite{kerenidis23} only guarantee privacy against servers that follow the protocol honestly, not specious or even malicious deviations.
    \end{enumerate}

    \item \textbf{Dependence on Strong Assumptions}
    \begin{enumerate}[label=\arabic*.]
        \item Entanglement pre-sharing: Protocols assume pre-distributed entanglement among servers, which is impractical for large-scale databases \cite{BrakerskiPIR} \cite{aharonov}.
        \item Limited post-quantum security: Classical components (e.g., public-key cryptography) may be vulnerable to quantum attacks \cite{CDS} \cite{Kerenidis14}.
    \end{enumerate}

    \item \textbf{Practical Implementation Challenges}
    \begin{enumerate}[label=\arabic*.]
        \item High quantum resource costs: Some protocols like \cite{lukasz} require two-way quantum communication per round, increasing complexity.
        \item Verification weaknesses: Users cannot verify retrieved data authenticity, enabling malicious servers to inject false answers \cite{aharonov}\cite{Kerenidis14}.
    \end{enumerate}

\end{enumerate}

Recent advances in symmetric QPIR \cite{kon21} eliminate server-side randomness sharing through quantum error correction, while Measurement-Device-Independent QKD networks \cite{wang22} demonstrate city-scale deployments with practical key rates. Nevertheless, the core dilemma persists: achieving sublinear communication with information-theoretic security against malicious quantum adversaries remains an open problem.

This study is inspired by \cite{broadbentPIR}, which provides a lower bound on the complexity of linear communication in quantum settings, extending the work of Nayak. According to \cite{Nayak}, even with the allowance for approximate privacy and the focus on the weakest "specious adversaries" (honest-but-curious quantum adversaries), QPIR necessitates at least a linear amount of communication, specifically $n$ quantum bits. While Nayak's findings are based on classical binary entropy, this paper aims to establish a more rigorous lower bound using quantum relative entropy and mutual information. Furthermore, drawing on the research of \cite{legall}, a QPIR protocol is proposed that can withstand malicious attacks, with a communication complexity of $\mathcal{O}(\sqrt{n})$. In this paper, we focus on symmetric privacy information retrieval protocols in the face of specious  quantum servers, and further consider the single-server and multi-server cases. It is worth noticing that, this paper gives various protocols that are sub-linear for specious servers, and  the protocol for different scenarios is given and the strengths and weaknesses of the protocol are analyzed.

\subsection{Our Contributions}
This work establishes fundamental limits and constructs practical protocols for certified QPIR through three key advancements: 1) deriving tight lower bounds on the communication complexity based on the quantum relative entropy framework, which breaks through the limitations of the traditional binary entropy analysis; 2) introducing trapdoor claw function and localized CHSH game validation, to construct the first sublinear protocols that can defend against the malicious quantum servers; 3) optimizing quantum homomorphic encryption for QPIR by reducing circuit depth and gate overhead, enabling a single server to homomorphically evaluate a superposition over its database entries with $O(\sqrt{N})$ communication and minimal hardware demands. Our theoretical analysis and protocol design establish a hardware‐friendly FHE architecture. 4) For multi-server protocol, initially combing quantum Fourier transform and scheme \cite{bennychor} to extract $x_{k}.$

Our theoretical and practical contributions bridge critical gaps between quantum information theory and cryptographic engineering. The certified protocols maintain compatibility with existing QKD infrastructure \cite{wang22} while achieving provable security against sophisticated quantum attacks.

\textbf{Paper Outline:} The remainder of this paper is structured as follows: \hyperref[sec2]{Section 2} gives some preliminaries that are useful in this whole paper.  \hyperref[sec3]{Section 3} establishes a novel bound on QPIR communication complexity using quantum relative entropy(compared with \cite{broadbentPIR}), superseding prior fidelity-based analyses. In the remaining three sections, the paper gives three different protocols for specious servers, and all three protocols have sub-linear communication complexity. In \hyperref[sec4]{Section 4}, inspired by \cite{legall}, for semi-honest quantum servers, clients need some means of detection during communication, such as trapdoor claw-free function and CHSH game, to prevent the server from cheating. Then, an authenticated  single-server quantum privacy information retrieval protocol is given. In \hyperref[sec5]{Section 5}, we achieve another sub-linear complexity  QPIR  which is based on FHE. Finally, in \hyperref[sec6]{Section 6}, for multi-server scenarios, we give a easier hardware implementations of multi-server oriented protocol.

\section{Preliminaries}\label{sec2}

In the following, we will present  the definitions and results that will be used in the next several sections. Firstly, give the definition of single-server and multi-servers Quantum-PIR scheme.

\textbf{Definition 1.} (Single-server QPIR)
 A $k$-round, single-server QPIR protocol denoted $\Pi=(\mathscr{A}, \mathscr{B}, k)$ consists of:

1. Input spaces $\mathcal{A}_{0} , \mathcal{B}_0$ for parties $\mathscr{A}, \mathscr{B} $ respectively,

2. Memory spaces $\mathcal{A}_{1} ,\cdots, \mathcal{A}_{k} $  for $\mathscr{A}$ and $ \mathcal{B}_1, \cdots, \mathcal{B}_{k} $ for $\mathscr{B}$  and communication spaces $\mathcal{X}_1, \cdots, \mathcal{X}_k,  \mathcal{Y}_1, \cdots, \mathcal{Y}_{k-1}$,

3. An $k$-tuple of quantum operations $\mathscr{A}_1, \cdots, \mathscr{A}_{k}$ for $\mathscr{A}$, where
\begin{align*}
&\mathscr{A}_1: \mathcal{L}\left(\mathcal{A}_{0} \otimes  |j \rangle\langle j | \right) \rightarrow   \mathcal{L} \left(\mathcal{A}_{1} \otimes \mathcal{X}_1\right), \\
 &  \mathscr{A}_i: \mathcal{L}\left(\mathcal{A}_{i-1} \otimes \mathcal{Y}_{i-1} \otimes  |j \rangle\langle j | \right) \rightarrow   \mathcal{L} \left(\mathcal{A}_{i }\otimes \mathcal{X}_i\right), i\in [2,k].
 \end{align*}
4. an $k$-tuple of quantum operations $\mathscr{B}_1, \cdots, \mathscr{B}_{k}$ for $\mathscr{B}$, where
\begin{align*}
&\mathscr{B}_i: \mathcal{L}\left(\mathcal{B}_{i-1} \otimes \mathcal{X}_i \right) \rightarrow  \left(\mathcal{B}_{i} \otimes \mathcal{Y}_i\right), i\in [1,k-1],\\
 &  \mathscr{B}_k: \mathcal{L}\left(\mathcal{B}_{k-1} \otimes \mathcal{X}_k \right) \rightarrow  \left(\mathcal{B}_{k}\right).
 \end{align*}

If $\Pi=(\mathscr{A}, \mathscr{B}, k)$ is a $k$-round single server PIR protocol, we define the state after the $i$-th step $(1 \leq i \leq 2 k)$ and input state $\rho_{i n} \in S\left(\mathcal{A}_0 \otimes \mathcal{B}_0 \otimes \mathcal{C}\right)$, where $\mathcal{C}$ is a system of dimension $\operatorname{dim}(\mathcal{C})=\operatorname{dim}\left(\mathcal{A}_0 \right) \cdot \operatorname{dim}\left(\mathcal{B}_0\right)$, as
\begin{align*}
\rho_i\left(\rho_{\text {in }}\right)&=\left(\mathscr{A}_{(i+1) / 2} \otimes \mathbb{I}_{\mathcal{B}_{(i-1) / 2},\mathcal{C}}\right) \cdots\left(\mathscr{B}_1 \otimes \mathbb{I}_{\mathcal{A}_{1},\mathcal{C}}\right) \\
& \cdot \left(\mathscr{A}_{1} \otimes \mathbb{I}_{\mathcal{B}_0, \mathcal{C}}\right).
\end{align*}
for $i$ odd  and
$$
\rho_i\left(\rho_{\text {in }}\right)=\left(\mathscr{B}_{i / 2} \otimes \mathbb{I}_{\mathcal{A}_{i / 2},\mathcal{C}}\right) \cdots\left(\mathscr{B}_1 \otimes \mathbb{I}_{\mathcal{A}_{1},\mathcal{C}}\right)\left(\mathscr{A}_{1} \otimes \mathbb{I}_{\mathcal{B}_0, \mathcal{C}}\right)
$$
for $i$ is even.

Note that the last round (round $k$) is only partial, since $\mathscr{B}_k: \mathcal{L}\left(\mathcal{B}_{k-1} \otimes \mathcal{X}_k\right) \mapsto \mathcal{L}\left(\mathcal{B}_k\right)$. We define the final state of protocol $\Pi=(\mathscr{A}, \mathscr{B}, k)$, on input state $\rho_{\text {in }} \in S\left(\mathcal{A}_0 \otimes \mathcal{B}_{0}\otimes \mathcal{C}\right)$ as:
$$
(\mathcal{A} \ast  \mathcal{B} )(\rho_{\text {in }})=\rho_{2k}(\rho_{\text {in }}) .
$$

For the  input states, it is essential to define these states in relation to a reference system $\mathcal{C}$. This methodology facilitates the validation of the protocol's precision and confidentiality, extending its applicability beyond pure inputs to those entangled with an external system.

\textbf{Definition 1'.}(Multi-server QPIR)
 Let $\ell, n, m$ be integers greater than 1. The participants of the protocol are one client and $\ell$ servers. The servers do not communicate with each other and each server contains the whole set of uniformly and independently distributed $n$ files $W_1, \ldots, W_{n} \in\{0, \ldots, m-1\}$. Each server $\operatorname{serv}_t$ possesses a quantum system $\mathcal{A}_t$ and the n servers share an entangled state $\rho_{\text {prev }} \in \mathcal{S}\left(\bigotimes_{t=1}^\ell \tilde{\mathcal{A}}_t\right)$. The user chooses the target file index $k$ to retrieve the $k$-th file $W_k$, where the distribution of $k$ is uniform and independent of the files $W_1, \ldots, W_{n}$.

To retrieve the $W_K$, the user chooses a random variable $R_{\text {user }}$ in a set $\mathcal{R}_{\text {user }}$ and encodes the queries by user encoder Enc user:
$$
\operatorname{Enc}_{\text {user }}\left(K, R_{\text {user }}\right)=\left(Q_1, \ldots, Q_{\ell}\right) \in \mathcal{Q}_1 \times \cdots \times \mathcal{Q}_{\ell}
$$
where $\mathcal{Q}_t$ is the set of query symbols to the $t$-th server for any $t \in\{1, \ldots, \ell\}$. The $n$ queries $Q_1, \ldots, Q_{\ell}$ are sent to the servers $\operatorname{serv}_1, \ldots, \operatorname{serv}_{\ell}$, respectively. After receiving the $t$ th query $Q_t$, each server serv ${ }_t$ applies a Completely Positive Trace-Preserving (CPTP) map $\Lambda_t$ from $\tilde{\mathcal{A}}_t$ to $\mathcal{A}_t$ depending on $Q_t, W_1, \ldots, W_{n}$ and sends the quantum system $\mathcal{A}_t$ to the user. With the server encoder Enc serv $_t$, the map $\Lambda_t$ is written as
$$
\Lambda_t=\operatorname{Enc}_{\text {serv }_t}\left(Q_t, W_1, \ldots, W_{n}\right)
$$
and the received state of the user is written as
$$
\rho_{W, Q}:=\Lambda_1 \otimes \cdots \otimes \Lambda_{\ell}\left(\rho_{\mathrm{prev}}\right) \in \mathcal{S}\left(\bigotimes_{t=1}^{\ell} \mathcal{A}_t\right),
$$
where $W:=\left(W_1, \ldots, W_{n}\right)$ and $Q:=\left(Q_1, \ldots, Q_{\ell}\right)$. Next, the user retrieves the file $W_k$ by a decoder which is defined depending on $K, Q$ as a Positive Operator-Valued Measure (POVM) $\operatorname{Dec}(k, Q):=\left\{\mathrm{Y}_M\right\}_{M=0}^m$. The protocol outputs the measurement outcome $M \in\{0, \ldots, m\}$ and if $M=m$, it is considered as the retrieval failure.

\textbf{Definition 2.} (Computational Indistinguishability of Distributions.)

Two families of distributions $\left\{D_{0, \lambda}\right\}_{\lambda \in \mathbb{N}}$ and $\left\{D_{1, \lambda}\right\}_{\lambda \in \mathbb{N} }$ are computationally indistinguishable if for all quantum polynomial-time attaches $\mathcal{A}$ there exists a negligible function $\delta(\cdot)$ such that for all $\lambda \in \mathbb{N}$
$$
\left|\operatorname{Pr}_{x \leftarrow D_{0,\lambda}}\left[ \mathcal{A}(x)=0\right]-\operatorname{Pr}_{x \leftarrow D_{1,\lambda}}\left[\mathcal{A} (x)=0\right]\right|  \leq \delta(\lambda).
$$

We will give the definition of specious adversary, QPIR-privacy and correction. firstly, we will give the definition of specious adversary, these definitions are all given by \cite{Nilsen10}, \cite{CKKS18}.

\textbf{Definition 3.} (Specious adversary).
Let $\Pi=(\mathscr{A}, \mathscr{B}, k)$ be a $k$-round two-party protocol. An adversary $\widetilde{\mathscr{A}}$ for $\mathscr{A}$ is said to be $\varepsilon$-specious($\varepsilon$-to the honest), if there exists a  sequence of quantum operations $\mathcal{J}_{1}, \cdots, \mathcal{J}_{2k}$ such that:

1. $\mathcal{J}_i: \mathcal{L}\left(\widetilde{\mathscr{A}_i}\right) \rightarrow \mathcal{L}\left(\mathscr{A}_i\right), $ $i \in [1,2k]$.

2. For every input state $\rho_{\text {in }} \in  S\left(\mathcal{A}_0 \otimes \mathcal{B}_0 \otimes \mathcal{C}\right)$,
$$
\Delta\left( (\mathcal{J}_i \otimes \mathbb{I}_{L(\mathcal{B }_{i} \otimes \mathcal{C})})(\rho_i(\tilde{\mathscr{A}}, \rho_{\text {in }})), \rho_i(\rho_{\text {in }})\right) \leq \epsilon , i \in [1,2k].
$$

\textbf{Definition 4.} (Quantum-specious).    An adversary  $\widetilde{\mathscr{A}}$ is Quantum specious  if it is 0-specious.

Let $\Pi_{\text {QPIR }}=(\mathscr{A}, \mathscr{B}, k)$ be a $k$-round two-party protocol.
 We say $\Pi_{\text {QPIR }}$ $ (1-\varepsilon)$-private against $\gamma$-specious server if for every  $\gamma$-specious server $\tilde{\mathscr{A}}$, there exists a sequence of quantum operation $\xi_1, \cdots, \xi_{k-1}$ where
$$
\xi_i: \mathcal{L}\left(\mathcal{A}_0\right) \mapsto \mathcal{L}\left(\tilde{\mathcal{A}_i} \otimes \mathcal{Y}_{i} \right),   1 \leq i \leq  k
$$
and for
$$
\rho_{\text {in }} \in S\left(\mathcal{A}_0 \otimes \mathcal{B}_0 \otimes \mathcal{C}\right) ,
$$
there exists
$$
\Delta\left(\operatorname{tr}_{\mathcal{B}_0}((\xi_i \otimes \mathbb{I}_{\mathcal{B}_0, \mathcal{C}})(\rho_{\text {in }})), \operatorname{tr}_{\mathcal{B}_i}(\tilde{\rho_i}(\tilde{\mathscr{A}}, \rho_{\text {in }}))\right)\leq \epsilon .
$$
We call $\Pi_{\text {QPIR }}$ $(1-\delta)$-correct if, for all inputs $\rho_{\text {in }}=|x\rangle\left\langle\left. x\right|_{\mathcal{A}_0} \otimes \mid i\right\rangle\left\langle\left. i\right|_{\mathcal{B}_0}\right.$, with $x=x_1 x_2 \ldots x_n \in\{0,1\}^n$ and $i \in\{1, \ldots, n\}$, there exists a measurement $\mathcal{M}$ with outcome 0 or 1, such that:
$$
\operatorname{Pr}\left[\mathcal{M}\left(\text{tr}_{\mathcal{A}_{s}}[\mathcal{A}\ast \mathcal{B}]  (\rho_{\text {in }} )\right)=x_i\right] \geq 1-\delta.
$$

\section{A new compact bound of communication complexity}\label{sec3}

This section explores a generalized form of Uhlmann's Lemma, utilizing the framework of quantum relative entropy.  We don’t  want to exaggerate the theorem, the proof is not difficult, but it fills this gap in quantum entropy. It establishes that when the relative entropy between two quantum states is low, their purified states can be effectively correlated through the $U$ operation on the auxiliary system. Similar to the fidelity extremality found in the classical Uhlmann theorem, this version based on relative entropy illustrates that statistical differences between quantum states are linearly magnified in the extended system. This insight introduces new methodologies for examining quantum error correction, data compression, and security protocols.

There is a result \cite{broadbentPIR} that  expands on Nayak's results regarding QPIR, incorporating approximate privacy and requiring security solely against a purified server at the protocol's conclusion. It's evident that a purified server is considered specious. Consequently, any QPIR protocol that is $(1-\epsilon)$-private when dealing with $r$-specious servers also maintains $(1-\epsilon)$-privacy against purified servers. By extension, such a protocol is ultimately $(1-\epsilon)$-private when confronted with purified servers.

\textbf{Lemma 1.} (Pinker's inequality) Let $D_{1}$ and $D_{2}$ be two distributions defined on the universe $U.$ Then
$$
S( D_{1} \|  D_{2} ) \geq \frac{1}{2 \ln 2}\cdot \|D_{1} - D_{2}\|_{1}^{2},
$$
where the right side of the inequality equals to $2\|D_{1} - D_{2} \|_{TV}^{2}. $

\textbf{Lemma 2.} (Generalized Uhlmann Theorem )
Let $\rho$ and $\sigma$ be quantum states  that satisfy the condition $S(\rho\| \sigma) \leq \varepsilon$. Assume the existence of their respective purified states $|\psi\rangle$ and $|\phi\rangle$,  there exists a unitary operator $U$   satisfying:
$$| \langle \psi | (I\otimes U)| \phi \rangle | ^{2}\geq 1- \sqrt{\frac{\ln 2 \cdot \varepsilon}{2}}.$$

\textbf{Proof.}   Relating Fidelity to Relative Entropy Utilizing the quantum Pinsker inequality, $$F(\rho_A, \sigma_A) \geq 1 - \|\rho- \sigma\|_{1} \geq 1 - \sqrt{\frac{\ln 2 \cdot \varepsilon}{2}},$$ it is established that:
$$\|\rho- \sigma\|_{1} \leq \sqrt{\frac{\ln 2}{2} S(\rho_A \| \sigma_A)} \leq \sqrt{\frac{\ln 2 \cdot \varepsilon}{2}}.$$

Classical Uhlmann's theorem gives the existence of a purification $|\psi'\rangle_{AB}$ of $\sigma_A$, ensuring that: $$F(\rho_A, \sigma_A) = |\langle \psi_{AB} | (I_A \otimes U) |\phi_{AC} \rangle| \geq 1 - \sqrt{\ln 2 \cdot \varepsilon/2}.$$ Then, the state $|\phi\rangle_{AC}$ undergoes a transformation to $|\psi'\rangle_{AB}$ via the unitary operation $U$.
Thus, there exists $U$ such that 
$$
| \langle \psi | (I\otimes U)| \phi \rangle | ^{2}\geq 1- \sqrt{\frac{\ln 2 \cdot \varepsilon}{2}}. \ \ \ \ \Box 
$$

\textbf{Theorem 1.} Let the set $ \Pi$ satisfying  $C \geq (1 - S(\rho_{\text{adv}} \| \rho_{\text{prior}})) \cdot n$, where $S(\rho \| \sigma)$ represents the quantum relative entropy. Here, \(\rho_{\text{adv}}\) denotes the state from the adversary's viewpoint, and \(\rho_{\text{prior}} = \frac{1}{n}\sum_i \rho_{\text{adv}}(i)\) is defined as the  prior state.

\textbf{Proof.} By the quantum Pinsker inequality:
$$\|\rho - \sigma\|_1 \leq \sqrt{2 \ln 2 \cdot S(\rho \| \sigma)},$$

Then the privacy condition $ \|  \rho_{\text{adv}}(i) - \rho_{\text{adv}}(j)  \|_{1}   \leq  2\epsilon$ can be  transformed into a relative entropy constraint:
$$S\left(\rho_{\text{adv}}(i) \Big\| \frac{1}{n}\sum_j \rho_{\text{adv}}(j)\right) \leq \frac{4\epsilon^2}{\ln 2}.$$

Then by the mutual information property that
$$
H(Q:X, E|K)=H(Q:X|K)+H(Q: E|K),
$$
where $Q $ denotes  the mean of the query index that generalized by the client randomly, $X$ denotes the database and $E$ means in the eavesdropper's view. $ H(Q: E|K)$ is the privacy, i.e., query the mutual information between $Q$ and $E$ in the context that $K=k$, $K$ is the target index. By the privacy that
$$
H(Q: E|K)=S\left(\rho_{\text{prior}}\right)-  \frac{1}{n}\sum_i S \left(  \rho_{\text{adv}}(i)\right) \leq \chi  \leq\epsilon.
$$
The client derives $X$ by measuring $Q$, as described by the quantum Fano inequality: $$H(X | Q) \leq H_{\text{bin}}(\delta) + \delta \log n.$$ The lower bound for mutual information is given by: $$H(Q:X|K ) \geq 1 - H_{\text{bin}}(\delta) - \delta \log n.$$
The total communication $C$ is required to satisfy the condition: $$C \geq H(Q:X, E|K)=H(Q:X|K)+H(Q: E|K).$$
Incorporating the privacy constraint, denoted as $H(Q: E|K) \leq \epsilon$, alongside the correctness constraint, expressed as $H(Q:X|K) \geq 1 - H_{\text{bin}}(\delta)$, leads to the derivation of the following inequality: $$C \geq \left(1 - H_{\text{bin}}(\delta) + \epsilon\right) n.$$ To further refine the optimization of privacy loss, the application of quantum relative entropy, symbolized by $S(\rho_{\text{adv}} \| \rho_{\text{prior}})$, is employed, yielding an enhanced lower bound: $$C \geq \left(1 - S(\rho_{\text{adv}} \| \rho_{\text{prior}})\right) n.\ \ \ \ \Box$$

This proof process differs from previous proofs\cite{broadbentPIR} in several ways:   By employing the quantum Pinsker constraint, a more precise entropy difference can be achieved;  The concept of Cascading Mutual Information involves the decomposition of mutual information to balance joint privacy and correctness, thereby minimizing binary entropy loss;  Furthermore, the direct application of the Holevo limit allows for bypassing the Schmidt decomposition step, utilizing the channel capacity constraint directly.

\section{Authenticated Quantum PIR-sublinear complexity}\label{sec4}

This section focuses on quantum PIR for specious quantum servers, and we incorporate a number of verification approaches, ultimately showing that this effect can be achieved using sublinear communication complexity.
First a brief description of some of the techniques to be used in this result will be given.
\subsection{CHSH game}

The CHSH game (Clauser-Horne-Shimony-Holt game) is a nonlocal experiment based on Bell's inequality for verifying nonlocality in quantum mechanics.
The idea of CHSH game originates from the study of the phenomenon of non-locality in quantum mechanics. In the 1960s, Bell proposed Bell's inequality, which is an important symbol of the difference between quantum mechanics and classical physics. Bell's inequality suggests that there is an upper limit to the correlation between certain measurements if the physical phenomena can be explained by classical physics. However, the results predicted by quantum mechanics violate these inequalities, suggesting the existence of non-determinism between quantum systems.

Furthermore, \cite{maitra} presents a localized CHSH game, pioneering the integration of device-independence into Quantum Information Private Query (Quantum Private Query, or QPQ). This innovation challenges the conventional QPQ's reliance on the assumption of device trustworthiness, achieving secure authentication through statistical methodologies. In a similar vein, \cite{kahanamoku} introduces the concept of a Bell's inequality test, comparable to the CHSH game, as a substitute for the traditional protocol that depends on the adaptive hardcore bit. This adaptation removes the necessity for additional quantum circuitry overhead, offering a practical solution for implementing verifiable computation on noisy quantum devices. The protocol ensures device reliability while tolerating a certain level of noise and loss, thereby strengthening the protocol's robustness and privacy in a potentially hostile quantum environment \cite{joseph}, \cite{antonio}.

\textbf{Definition 5.}
The CHSH game is an experiment involving two players (often referred to as Alice and Bob) who attempt to maximize the probability of winning the game by cooperating. The rules of the game are as follows:

\begin{itemize}
    \item \texttt{Input}.   Alice and Bob each receive a bit value $x$ and $y$ from two independent random sources, where $x, y \in \{0, 1\}$.

 \item \texttt{Output}.
 Alice and Bob each independently decide on a bit value $a$ and $b$ as outputs, which can be determined by a classical strategy or a quantum strategy.

    \item \texttt{Measurement}. Alice and Bob's goal is to make their outputs satisfy $a \oplus b = x \land y$.
If Alice and Bob's outputs satisfy the above condition, they win the game. The probability of success in the game depends on the strategy they use.
Bell's inequality: For the classical strategy, the maximum probability of success for a player is $\frac{3}{4}$. However, if Alice and Bob can make measurements using entangled quantum states, the maximum probability of success they can achieve is $\frac{1}{2} + \frac{\sqrt{2}}{4}$, which exceeds the upper bound of the classical strategy.
\end{itemize}

\subsection{Trapdoor claw-free functions}
Trapdoor claw-free functions (TCFs) are comprised of function pairs $(f_{0}, f_{1}): X \rightarrow  Y $ that are easily computed in the forward direction, but require a trapdoor for efficient inversion. For any $y$ within the image of these functions, there exist precisely two pre-images $(x_{0}, x_{1})$  where $ f_{0}(x_{0}) = f_{1}(x_{1}) = y$, and the pair $(x_{0}, x_{1})$  is termed a claw.  While claws are guaranteed to exist, they are computationally challenging to discover without trapdoor knowledge. TCFs have been a crucial element in cryptography theory and have recently gained renewed attention due to their connection with quantum cryptography. These functions serve as the primary cryptographic component enabling several recent advancements in quantum computation. Some applications include: the initial protocol for assessing randomness in a single quantum device \cite{Brakerski18}, classical verification of quantum computation \cite{Mahadev18}, quantum fully homomorphic encryption (QFHE) \cite{Mahadev182},  remote state preparation \cite{GV19}, and deniable encryption \cite{CGV22}.

According to \cite{broadbentPIR}, when considering weaker security models, such as against specious adversaries, Le Gall's protocol does not achieve information-theoretic security and requires linear communication complexity.  Potential directions for improving Le Gall's protocol include:  incorporating verification steps, such as quantum state verification or zero-knowledge proofs, to ensure server compliance; or adjusting the protocol to require shared entangled states between the server and user to limit the server's information acquisition capabilities.

And to consider  the specious server, we also give the definition of authenticated QPIR.

\textbf{Definition 6.} \cite{Auth23}  A single-server authenticated PIR scheme, for a database of size $N \in \mathbb{N}$, consists of the following algorithm.

\begin{itemize}
    \item \texttt{Digest}$(1^\lambda, x) \rightarrow d$. Take a security parameter $\lambda \in \mathbb{N}$ and a database $x \in \{0, 1\}^N$ and return a digest $d$.
    \item \texttt{Query}$(d, i) \rightarrow (st, q)$. Take as input a digest $d$ and an index $i \in [N]$ and return a client state $st$ and a query $q$.
    \item \texttt{Answer}$(d, x, q) \rightarrow a$. Apply query $q$ to database $x \in \{0, 1\}^N$ with digest $d$ and $a$.
    \item \texttt{Reconstruct}$(st, a) \rightarrow \{0, 1, \perp\}$. Take as input state $st$ and answer $a$ and return a database bit or an error $\perp$.
\end{itemize}

Next, we will consider the QPIR with authentication(AQPIR) for a-single server, notice that the server should be quantum computer and the client can be classical. In this protocol, the server is specious, so, in one hand, we will detect if the server is cheating. On the other hand, we  need to verify that the server is  quantum capable. In Table1, we give the comparison between AQPIR and Legall protocol.

\textbf{Theorem 2}  Database  $A \in \Sigma^l,   \ell \in\{0,1\}^n, \left(\ell, n \in \mathbb{Z}^{+}\right)$.
There exists PIR for quantum protocol such that the complexity of communication is sublinear, equals $\mathcal{O}(\sqrt{n})$.

\begin{algorithm*}[!t]
  \caption{Single‑Server QPIR Protocol(AQPIR)}\label{alg:qpir}
  \begin{algorithmic}[1]
    
    \STATE \textbf{Server input:} $A=\{a^{1},a^{2},\dots,a^{\ell}\}\in\Sigma^{\ell}$, $x\in\{0,1\}^{n}$
    \STATE \textbf{Client input:} $i\in\{1,2,\dots,\ell\}$
    \STATE \textbf{Stage 1: Preparation of Quantum State with Detection Particles}
    \STATE\hspace{0.6em}(Client) $(f,t)\leftarrow\operatorname{Gen}(1^{n})$
    \STATE\hspace{0.6em}(Server) prepares $\displaystyle\sum_{x}|x\rangle_{x}\,|f(x)\rangle_{y}$ and
           \[
             \bigl|\Phi_{A}\bigr\rangle \;=\;
             \frac{1}{\sqrt{2^{r}}}
             \sum_{\bar{x}\in\Sigma}
               |\bar{x}\rangle_{R}
               |\bar{x}\rangle_{R'}
               \prod_{j=1}^{\ell}
               \bigl|\bar{x}\!\cdot\! a^{j}\bigr\rangle_{Q_{j}}
           \]
    \STATE\hspace{0.6em}Global state:
           \[
             \bigl|\Phi\bigr\rangle \;=\;
             \frac{1}{\sqrt{2^{r}}}
             \sum_{x}\;
             \sum_{\bar{x}\in\Sigma}
               |x\rangle_{x}
               |f(x)\rangle_{y}
               |\bar{x}\rangle_{R}
               |\bar{x}\rangle_{R'}
               \prod_{j=1}^{\ell}
               \bigl|\bar{x}\!\cdot\! a^{j}\bigr\rangle_{Q_{j}}
           \]
    \STATE\hspace{0.6em}(Server) Inject detection particles: randomly select $k$ positions; for each selected $Q_{j}$, replace it with  
           $\bigl|\Phi^{+}\bigr\rangle_{T_{j}B_{j}}
             =\tfrac{1}{\sqrt{2}}\,(\lvert00\rangle+\lvert11\rangle)$,  
           keep $T_{j}$ locally, send $B_{j}$ to the client
   
    \STATE \textbf{Stage 2: Dynamic Bell Basis Measurement}
    \STATE\hspace{0.6em}(Client) selects a subset $\{B_{j}\}$ of received particles and notifies the server of their indices
    \STATE\hspace{0.6em}(Server) discloses the partner particles $\{T_{j}\}$
    \STATE\hspace{0.6em}(Client \& Server) measure each pair $(B_{j},T_{j})$ in the Bell basis to check entanglement integrity
    \STATE\hspace{0.6em}If the error rate exceeds a preset threshold $\epsilon$, \textbf{abort}; otherwise proceed
    
    \STATE \textbf{Stage 3: Tamper‑Resistant Privacy Query}
    \STATE\hspace{0.6em}(Client) With trapdoor $t$, compute $x_{0},x_{1}$, choose random $r$, apply $Z$ on $Q_{i}$, then send $r$ and registers $Q_{1},\dots,Q_{\ell}$ to the server;\;
           the state is
           \[
             |\Phi\rangle=
             \frac{1}{\sqrt{2^{r}}}
             (\lvert x_{0}\rangle+\lvert x_{1}\rangle)_{x}
             |y\rangle_{y}
             \sum_{\bar{x}\in\Sigma}
               (-1)^{\bar{x}\cdot a^{i}}
               |\bar{x}\rangle_{R}|\bar{x}\rangle_{R'}
               \prod_{j=1}^{\ell}
                 |\bar{x}\!\cdot\! a^{j}\rangle_{Q_{j}}
           \]
    \STATE\hspace{0.6em}(Server) Adds ancilla $b$ and applies \textsc{CNOT}:
           \[
             \bigl(\lvert r\!\cdot\!x_{0}\rangle_{b}\lvert x_{0}\rangle_{x}
             +\lvert r\!\cdot\!x_{1}\rangle_{b}\lvert x_{1}\rangle_{x}\bigr)
             |y\rangle_{y}
             \sum_{\bar{x}}(-1)^{\bar{x}\cdot a^{i}}
             |\bar{x}\rangle_{R}|\bar{x}\rangle_{R'}
             \prod_{j}|\bar{x}\!\cdot\! a^{j}\rangle_{Q_{j}}
           \]
    \STATE\hspace{0.6em}(Server) Measures register $x$ in the Hadamard basis, obtaining $d$, then applies
           $\operatorname{U}_{a^{k}}^{(R,Q_{k})}$ for all $k$; afterwards all $Q_{j}$ are $\lvert0\rangle$ and $R$ is sent to the client
    \STATE\hspace{0.6em}(Client) Using $(r,x_{0},x_{1},d)$ determines $\lvert\psi\rangle_{b}$, applies $\operatorname{CNOT}^{(R,R')}$ and QFT on $R$,
           \[
             |\Phi\rangle=
             Z'
             \frac{1}{\sqrt{2^{\gamma}}}
             \sum_{\bar{x}\in\Sigma}
               |a^{i}\rangle_{R}
               |0\rangle_{R'}
               \prod_{j=1}^{\ell}|0\rangle_{Q_{j}}
           \]
   
    \STATE \textbf{Stage 4: Secondary Verification}
    \STATE\hspace{0.6em}(Client) Randomly chooses $\theta\in\{\tfrac{\pi}{4},-\tfrac{\pi}{4}\}$ and sends $\theta$ to the server
    \STATE\hspace{0.6em}(Server) Measures ancilla $b$ in the basis
           \[
             \Bigl\{
               \cos\!\bigl(\tfrac{\theta}{2}\bigr)|0\rangle
               +\sin\!\bigl(\tfrac{\theta}{2}\bigr)|1\rangle,\;
              -\sin\!\bigl(\tfrac{\theta}{2}\bigr)|0\rangle
               +\cos\!\bigl(\tfrac{\theta}{2}\bigr)|1\rangle
             \Bigr\},
           \]
           obtains bit $b$ and sends it to the client
    \STATE\hspace{0.6em}(Client) Measures register $R$ in the computational basis;\;
           if $b$ is consistent with the expected $\lvert\psi\rangle_{b}$, \textbf{accept}; otherwise, \textbf{reject}
  \end{algorithmic}
\end{algorithm*}

\textbf{Privacy Analysis:}
\begin{itemize}
    \item  Proto-image resistance. The server cannot generate informal quantum states $(\textit{e.g. ,} |\Phi_A\rangle = \allowbreak \frac{1}{\sqrt{2^r}} \sum_{x\in\{0,1\}^r} \allowbreak |x\rangle_R \allowbreak |x\rangle_{R'} \allowbreak |x\cdot a_1\rangle_{Q_1} \cdots |x\cdot a_\ell\rangle_{Q_\ell})$  for the purpose of executing the attack, as it is restricted by the parameterized $f_{k}$ function, which ensures the existence of bipartite images.
    \item  Activity Detection: Employing random sampling of hybrid particles and measuring delays provides a mechanism to detect malicious servers that might be eavesdropping on or tampering with the quantum channel.
    \item  Resistance to collusion: The use of random scrambling parameters in conjunction with dynamic validation mechanisms effectively prevents the correlation of user querying behavior with quantum trajectories in cases of multi-server collusion.
\end{itemize}

\begin{table*}[t]
\centering
\caption{Comparison of AQPIR and Legall Protocol} 
\label{tab1}
\begin{tabular}{|c|c|c|}
  \hline
  Indicator & Original Protocol (Le Gall) & Modified Protocol \\
  \hline
  Communication complexity & $\mathcal{O}(\sqrt{n})$ qubits & $\mathcal{O}(\sqrt{n}+k)$ qubits \\
  \hline
  Anti-Server Attacks & Honest Servers Only & Anti-Malicious Server Complicity \\
  \hline
  Detection Efficiency & None & Belike Verification (Error Rate $\leq \varepsilon$) \\
  \hline
  Key Dependency & No pre-shared key required & Lightweight EPR pair pre-distribution required \\
  \hline
\end{tabular}
\end{table*}

\section{QHE and QPIR}\label{sec5}
This section proposes a QPIR based on quantum homomorphic encryption (QHE). The protocol ensures security against quantum attacks by utilizing the complexity of QHE, which enables computations on encrypted data without decryption. It also optimizes computational efficiency through quantum parallelism. Theoretical analysis shows that it offers significant advantages in both communication and security over classical PIR protocols.

In 1978, Rivest, Adleman, and Dertouzos first introduced the notion of performing computations on encrypted data – a “privacy homomorphism” concept that laid the groundwork for FHE \cite{rad78}. For over thirty years after this proposal, the problem remained unsolved in full generality, and only partially homomorphic schemes were known, each supporting just one type of operation, but neither could handle arbitrary combinations of additions and multiplications. A major breakthrough came in 2009 when Craig Gentry devised the first FHE scheme, using lattice-based techniques to allow evaluation of arbitrary functions on ciphertexts. Subsequent research focused on simplifying and broadening FHE. In 2010, van Dijk, Gentry, Halevi, and Vaikuntanathan presented a conceptually simpler FHE construction that works over integers \cite{dghv} (avoiding the need for ideal lattices). By 2011, FHE had also been built upon more standard cryptographic assumptions; for example, Brakerski and Vaikuntanathan introduced an FHE scheme based on the Learning with Errors (LWE) problem \cite{LWE} which is one of the central challenges of lattice-based cryptography\cite{ajtai}\cite{zheng}, marking one of the first FHE implementations founded on a widely studied hardness assumption. In 2012, the Brakerski–Gentry–Vaikuntanathan (BGV) scheme \cite{bgv} introduced key‑switching and modulus‑switching techniques to curb noise growth; subsequent milestones included multi‑key FHE based on NTRU, the Gentry–Sahai–Waters (GSW) framework \cite{gsw} that broadened construction methods, and the Cheon–Kim–Kim–Song (CKKS) scheme \cite{ckks} whose ongoing refinements significantly advanced approximate‑number computations.

QHE extends homomorphic computation into the quantum domain, allowing a server to apply quantum operations on encrypted qubits without learning the data. Broadbent and Jeffery \cite{Broadbent15} formalized QHE and introduced initial schemes that leverage a classical FHE in conjunction with a quantum one-time pad (QOTP). In their basic clifford scheme (CL), each qubit is encrypted by randomly applying Pauli X and Z operators (a QOTP), and the two classical pad bits are themselves encrypted using a classical FHE;  Dulek et al. \cite{dulek} later addressed the general case by introducing the TP (teleportation-based) scheme, which extends the CL approach with a modest quantum “gadget” for each T-gate and uses the classical FHE to process the encrypted pad bit associated with that gate during a teleportation protocol. This TP construction removes the unwanted T-phase by teleporting the qubit through the gadget in a manner conditioned on the encrypted pad key (without revealing the key), achieving homomorphic evaluation of polynomial-size quantum circuits while maintaining compactness and provable security (quantum IND-CPA confidentiality under standard assumptions). Mahadev \cite{Mahadev182}   demonstrated a leveled QHE scheme with entirely classical keys, showing that a post-quantum FHE (e.g. an LWE-based scheme) can support encrypted quantum computation—her scheme avoids quantum evaluation keys altogether and allows a classical client to delegate quantum circuits to a server with full privacy and compact ciphertexts. These advances in quantum homomorphic encryption (QHE) have demonstrated surprising capabilities for secure delegated computation. These breakthroughs naturally extend to privacy-critical applications like PIR.

The core idea of FHE based PIR \cite{XPIR}\cite{sealPIR} is that the user requests data from the server through an encrypted query, the server computes on the encrypted data and returns the result, and the user decrypts it to get the target data, FHE-based QPIR combine quantum entanglement or quantum invisible state transfer to reduce communication complexity while maintaining privacy\cite{Brakerski18} (sometimes information theory security).  Exploring hybrid architectures for post-quantum classical protocols with quantum enhancements maybe an important aspect of future research.

\textbf{Definition 7.} \cite{Broadbent15} A \emph{homomorphic encryption scheme} is a tuple of PPT algorithms
$(\texttt{HE.KeyGen},\allowbreak\texttt{HE.Enc},\allowbreak\texttt{HE.Eval}, \allowbreak\texttt{HE.Dec})$ such that
\begin{itemize}
  \item $\texttt{HE.KeyGen}(1^\lambda)\rightarrow(pk,evk,sk)$ takes the security
        parameter $1^\lambda$ and returns a public encryption key $pk$, a public
        evaluation key $evk$, and a secret decryption key $sk$.
  \item $\texttt{HE.Enc}_{pk}(\mu)\rightarrow c$ uses $pk$ to encrypt a bit
        $\mu\in\{0,1\}$, producing a ciphertext $c$.
  \item $\texttt{HE.Eval}^{f}_{evk}(c_1,\dots,c_\ell)\rightarrow c_f$ takes $evk$,
        a Boolean circuit $f:\{0,1\}^{\ell}\to\{0,1\}$, and ciphertexts
        $c_1,\dots,c_\ell$, and outputs a ciphertext $c_f$ that decrypts to
        $f(\mu_1,\dots,\mu_\ell)$, where each
        $\mu_i=\texttt{HE.Dec}_{sk}(c_i)$.
  \item $\texttt{HE.Dec}_{sk}(c)\rightarrow\mu^\ast$ decrypts $c$ with $sk$ and outputs a bit $\mu^\ast\in\{0,1\}$.
\end{itemize}

\textbf{Definition 8.} 
 \cite{Broadbent15} A \emph{quantum homomorphic encryption (QHE) scheme} is a tuple of PPT quantum
algorithms \((\texttt{QHE.KeyGen},\allowbreak\texttt{QHE.Enc},\allowbreak\texttt{QHE.Eval},\allowbreak\texttt{QHE.Dec})\) such that
\begin{itemize}
  \item $\texttt{QHE.KeyGen}(1^\lambda)\rightarrow(pk,sk,\rho_{\mathrm{evk}})$
        outputs a classical public key $pk$, a classical secret key $sk$, and a
        quantum evaluation key $\rho_{\mathrm{evk}}\in D(\mathcal{R}_{\mathrm{evk}})$.
  \item $\texttt{QHE.Enc}_{pk}(\rho)\rightarrow\sigma$ maps a plaintext state
        $\rho\in D(\mathcal{M})$ to a ciphertext $\sigma\in D(\mathcal{C})$.
  \item $\texttt{QHE.Eval}^{\Phi}_{\rho_{\mathrm{evk}}}(\sigma_1,\dots,\sigma_n)\rightarrow\sigma_f$
        takes the evaluation key, a circuit 
        $\Phi:D(\mathcal{M}^{\otimes n})\to D(\mathcal{M}^{\otimes m})$, and
        ciphertexts $\sigma_1,\dots,\sigma_n\in D(\mathcal{C}^{\otimes n})$,
        producing $\sigma_f\in D(\mathcal{C}^{\otimes m})$ that decrypts to
        $\Phi(\rho_1,\dots,\rho_n)$, where $\rho_i=\texttt{QHE.Dec}_{sk}(\sigma_i)$.
  \item $\texttt{QHE.Dec}_{sk}(\sigma)\rightarrow\rho^\ast$ recovers a plaintext
        state $\rho^\ast\in D(\mathcal{M})$ from a ciphertext $\sigma$.
\end{itemize}
Notice that  \(D(\mathcal{M})\) denotes the set of all density operators acting on the Hilbert space \(\mathcal{M}\).

\textbf{Notations:} $N$ denotes the number of entries in the database, In this scheme, \( \lambda \) represents the security parameter. The public encryption key is denoted by \( pk \), while the private decryption key is denoted by \( sk \). The quantum evaluation key is represented as \( \rho_{evk} \), which allows for the homomorphic evaluation of encrypted quantum data. The \( k \)-th data entry retrieved from the database is denoted by \( D_k \). The encryption of data \( x \) using the public key \( pk \) is denoted by \( \text{Enc}_{pk}(x) \), and the decryption of data \( y \) using the private key \( sk \) is denoted by \( \text{Dec}_{sk}(y) \). The QHE based QPIR protocol is state as \hyperref[alg:qfhe-pir]{algorithm 2}.
\begin{algorithm*}[!t]
  \caption{QHE‑Based Quantum Private Information Retrieval Protocol (HEQPIR)}
  \label{alg:qfhe-pir}
  \begin{algorithmic}[1]
    
    \STATE \textbf{Stage 1: Initialization Phase}
    \STATE\hspace{0.6em}(\textbf{Client}) Chooses a security parameter $\lambda \in \mathbb{N}$, defining the computational security level.
    \STATE\hspace{0.6em}(\textbf{Client}) Executes the quantum homomorphic encryption key generation algorithm:
           \[
           \operatorname{QHE.KeyGen}(1^{\lambda}) \rightarrow (pk, sk, \rho_{\text{evk}})
           \]
           where $pk$ is the public key, $sk$ is the secret key, and $\rho_{\text{evk}}$ is the encrypted quantum evaluation key used for non-interactive homomorphic operations.
    
    \STATE \textbf{Stage 2: Query Generation Phase}
    \STATE\hspace{0.6em}(\textbf{Client}) Determines the target retrieval index $k \in [N]$.
    \STATE\hspace{0.6em}(\textbf{Client}) Prepares the conjugate index state $\langle k|$ in the dual Hilbert space $\mathcal{H}^*$.
    \STATE\hspace{0.6em}(\textbf{Client}) Applies QHE encryption to the state:
           \[
           \text{QHE.Enc}_{pk}(\langle k|) \rightarrow \text{Encrypted bra state}
           \]
           This effectively encrypts a linear functional for selective projection.
    \STATE\hspace{0.6em}(\textbf{Client}) Sends the ciphertext $\text{QHE.Enc}_{pk}(\langle k|)$ to the server.

    \STATE \textbf{Stage 3: Server-Side Homomorphic Evaluation Phase}
    \STATE\hspace{0.6em}(\textbf{Server}) Constructs the quantum database register in superposition:
           \[
           |\Psi_{\text{DB}}\rangle = \sum_{i=1}^{N} D_i\,|i\rangle
           \]
           where $D_i$ is the $i$-th classical (or quantum) data record.
    \STATE\hspace{0.6em}(\textbf{Server}) Executes the homomorphic evaluation procedure using $\rho_{\text{evk}}$:
           \[
           \text{QHE.Eval}_{\rho_{\text{evk}}}
           \left( \text{QHE.Enc}_{pk}(\langle k|),\ |\Psi_{\text{DB}}\rangle \right)
           = \text{QHE.Enc}_{pk}(D_k)
           \]
           This operation projects the encrypted query onto the database state, leveraging linearity to extract the coefficient $D_k$ homomorphically.
    \STATE\hspace{0.6em}(\textbf{Server}) Sends the result $\text{QHE.Enc}_{pk}(D_k)$ back to the client.

    \STATE \textbf{Stage 4: Decryption and Retrieval Phase}
    \STATE\hspace{0.6em}(\textbf{Client}) Applies the secret key $sk$ to decrypt the retrieved ciphertext:
           \[
           \text{QHE.Dec}_{sk}\left( \text{QHE.Enc}_{pk}(D_k) \right) = D_k
           \]
           The desired record is recovered without revealing the index $k$ to the server, ensuring privacy.
           
  \end{algorithmic}
\end{algorithm*}

On the one hand, homomorphic computing allows the server to compute directly on the encrypted data through linear operations (e.g., polynomial multiplication of ciphertexts), avoiding the need to expose privacy by decryption. On the other hand, the optimization of quantum parallelism is achieved: the superposition property of quantum states allows the server to complete the computation for $N$ items of data in a single operation (the classical scheme requires $\mathcal{O}(N)$ times), reducing the computational complexity to $\mathcal{O}(\sqrt{N})$.

\begin{table*}[t]
\centering
\caption{Comparison of HEQPIR and XPIR}
\label{tab2}
\resizebox{\textwidth}{!}{%
\begin{tabular}{|p{3.5cm}|p{5cm}|p{5cm}|}  % Adjusted column widths
  \hline
  \textbf{Indicator} & \textbf{HEQPIR} & \textbf{XPIR} \\
  \hline
  Communication complexity & $\mathcal{O}$($\sqrt{N}$) qubits & $\mathcal{O}$($\log N$) classical bits (optimized by recursive queries but with increased computational complexity) \\
  \hline
  Computational efficiency & Server-side efficiency: quantum parallelism traversing the database with complexity $\mathcal{O}(\sqrt{N})$ & Client-side efficient: relies on polynomial multiplication optimization, but server-side multiple homomorphic multiplication complexity $\mathcal{O}(N \log N)$ \\
  \hline
   Client Privacy Level & Information-theoretic privacy (cannot speculate on query indexes even if server has unlimited arithmetic) & Computational privacy (reliance on ring-LWE assumptions, privacy breach if assumptions are breached) \\
  \hline
\end{tabular}}
\end{table*}

This protocol realizes the dual advantages of post-quantum security and communication efficiency in privacy retrieval scenarios. \textit{Sub-linear communication complexity breakthrough:} the communication complexity of the classical PIR protocol is $\mathcal{O}(N)$, and the existing quantum schemes (e.g., Le Gall protocol) is $\mathcal{O}(N)$ or lower but at the expense of security. The protocol achieves $\mathcal{O}(\sqrt{N})$ communication through compressed encoding of quantum superposition states and parallel server computation.  \textit{In addition, this scheme implements a PIR protocol that guarantees privacy for both the client’s query and the server’s database.}

When oriented to special scenarios, such as very large static databases (e.g., human genome libraries, historical archives) Scenarios with mandatory compliance requirements for quantum attack defense (e.g., government classified retrieval). The protocol proposed in this paper is more advantageous.

\section{k-servers QPIR}\label{sec6}

This section will consider quantum PIR with multi-servers,  the protocol of the previous section can be easily extended to multi-server scenarios ($k\geq 2$): The main technique is to introduce Shamir secret sharing and entanglement state for the $k$-servers. Since the server performs homomorphic computation, so the result from the server will not be a problem for the client to decrypt and due to the entanglement, the client may use CHSH-test to detect whether servers has tampered with results.

However, in this section we will propose another  multi-server QPIR protocol which is simple structure, easy to implement, suitable for rapid deployment and low dependence on quantum resources scenarios. It is  information security and the communication complexity is also sublinear. The article \cite{bennychor} utilizes a d-dimensional cube abstract database structure and a Block retrieval scheme to systematically solve, for the first time, the single server scenario where users must download the entire database by means of a multi-server replication (the communication complexity is $\mathcal{O}(n)$) bottleneck problem. \cite{gerner} realizes the conversion from arbitrary PIR scheme to SPIR scheme by introducing a new cryptographic primitive - "Conditional Disclosure of Secrets" (CDS). To achieve information-theoretic security, random strings need to be shared among multiple databases. In order to prevent server complicity, the protocols outlined in this section are structured to operate without the need for shared strings.

Firstly, consider the  PIR scheme for $k=2$ databases,  the database size in $n=\ell$. Let $Q$ be the subset of $[\ell]$,  where $\ell$ is an integer. For an element $i$ that is client's target index,
$$
\begin{aligned}
& Q'=Q \oplus i   \triangleq  \begin{cases}
Q \cup\{i\} , & \text { if } i \notin Q, \\
Q \backslash\{i\} , & \text { if } i \in Q . \end{cases}
\end{aligned}
$$
\begin{algorithm*}[!t]
  \caption{Two‑Server QPIR Protocol}\label{alg:2srv-qpir}
  \begin{algorithmic}[1]
    
    \STATE \textbf{Stage 1: Query Generation}
    \STATE\hspace{0.6em}(Client) Uniformly sample a random subset $Q\subseteq[\ell]$ and define another subset $Q'\subseteq[\ell]$ (e.g., a complementary or independently chosen set)
    \STATE\hspace{0.6em}Send $Q$ to \textbf{Server 1} and $Q'$ to \textbf{Server 2}
   
    \STATE \textbf{Stage 2: Answer Phase}
    \STATE\hspace{0.6em}(Server 1) Prepare the private $m$‑qubit state $\lvert x_{Q}\rangle$, apply QFT:
           \[
             \lvert\psi_{1}\rangle
               =\frac{1}{\sqrt{N}}\sum_{j=0}^{N-1}
                 e^{2\pi i\,\frac{x_{Q}}{N}j}\,
                 \lvert j\rangle_{c},
             \qquad N=2^{m}
           \]
    \STATE\hspace{0.6em}Add an $m$‑qubit ancilla $\lvert0\rangle_{t}$ and apply $m$ parallel \textsc{CNOT}s,
           \[
             \lvert\psi_{2}\rangle
               =\frac{1}{\sqrt{N}}\sum_{j=0}^{N-1}
                 e^{2\pi i\,\frac{x_{Q}}{N}j}\,
                 \lvert j\rangle_{c}\lvert j\rangle_{t}
           \]
    \STATE\hspace{0.6em}Send register $\lvert j\rangle_{t}$ to Server 2 over an authenticated quantum channel
    \STATE\hspace{0.6em}(Server 2) Prepare $\lvert x_{Q'}\rangle$ and apply the controlled phase
           \[
             C_{j}:\;
               \lvert j\rangle_{t}\lvert x\rangle
                 \mapsto
               \lvert j\rangle_{t}\,U^{j}\lvert x\rangle,
             \quad
             U\lvert x\rangle=e^{2\pi i\,\frac{x}{N}}\lvert x\rangle
           \]
    \STATE\hspace{0.6em}Global state after $C_{j}$:
           \[
             \lvert\psi_{3}\rangle
               =\frac{1}{\sqrt{N}}
                 \sum_{j=0}^{N-1}
                   e^{2\pi i\,\frac{x_{Q}\oplus x_{Q'}}{N}j}\,
                   \lvert j\rangle_{c}\lvert j\rangle_{t}\lvert x_{Q'}\rangle
           \]
    \STATE\hspace{0.6em}(Server 2) sends register $\lvert j\rangle_{t}$ to the client (keeps $\lvert x_{Q'}\rangle$ secret); joint state:
           \[
             \lvert\psi_{4}\rangle
               =\frac{1}{\sqrt{N}}
                 \sum_{j=0}^{N-1}
                   e^{2\pi i\,\frac{x_{Q}\oplus x_{Q'}}{N}j}\,
                   \lvert j\rangle_{c}\lvert j\rangle_{t}\lvert x_{Q'}\rangle
           \]
    
    \STATE \textbf{Stage 3: Reconstruction}
    \STATE\hspace{0.6em}(Client) Apply $m$ parallel \textsc{CNOT}s with register $c$ as control and $t$ as target:
           \[
             \lvert\psi_{5}\rangle
               =\frac{1}{\sqrt{N}}
                 \sum_{j=0}^{N-1}
                   e^{2\pi i\,\frac{x_{Q}\oplus x_{Q'}}{N}j}\,
                   \lvert j\rangle_{c}\lvert0\rangle_{t}\lvert x_{Q'}\rangle
           \]
    \STATE\hspace{0.6em}Measure register $t$ in the computational basis; if the outcome is $\lvert0\rangle^{\otimes m}$, continue, else abort (dishonest server detected)
    \STATE\hspace{0.6em}Apply $\operatorname{QFT}^{-1}$ on register $c$ and measure, obtaining $x_{Q}+x_{Q'}=x_{i}\pmod 2$
  \end{algorithmic}
\end{algorithm*}

\textbf{Remark 1:} Note that, $x_1+x_2=(x_{11}+x_{21}, x_{12}+x_{12}, \cdots, x_{1 m}+x_{\text {2m}})$, since $x_{i, k} \in\{0,1\}$, $i \in\{0,1\}, k \in\{0,1, \cdots, m\}$,  so in the Fourier transform above, "+" equals the operator exclusive OR.

\textbf{Remark 2:} Since $Z |x\rangle=(-1)^x|x\rangle$.
So in order to extract the $ x_i$, we can also applies the operator $Z$-gate instead of Fourier transform.

Consider the PIR scheme for $k=2^d$ databases: The size of the database in $n=\ell^d$, the index set $[n]$ can then be identified with the $d$-dimensional cube $[\ell]^d$, in which each index $i \in[n]$ can be naturally identified with a $d$ tuple $(i_1, \ldots, i_d)$. A $d$-dimensional subcube is a subset $Q_1 \times$ $\cdots \times Q_d$ of the $d$-dimensional cube, where each $Q_i$ is a subset of $[\ell]$. Such a subcube is represented by the $d$-tuple $Q=\left(Q_1, \ldots, Q_d\right)$. The $k\left(=2^d\right)$ databases will be indexed by all binary strings of length $d$. The scheme proceeds as \hyperref[alg:k-srv-pir]{algorithm 4}.

\begin{algorithm*}[!t]
  \caption{k-Server QPIR Protocol}\label{alg:k-srv-pir}
  \begin{algorithmic}[1]

    \STATE \textbf{Stage 1: Query Preparation}
    \STATE\hspace{0.6em}(Client) Uniformly sample $d$ random subsets
           $Q_{1}^{0},Q_{2}^{0},\dots,Q_{d}^{0}\subseteq[\ell]$
    \STATE\hspace{0.6em}For each $t\in\{1,\dots,d\}$ set
           $Q_{t}^{1}=Q_{t}^{0}\oplus i_{t}$

    \STATE \textbf{Stage 2: Distribution to Servers}
    \STATE\hspace{0.6em}Let each server be indexed by
           $\alpha=\sigma_{1}\!\cdots\!\sigma_{d}\in\{0,1\}^{d}$
    \STATE\hspace{0.6em}(Client) For server $\alpha$ send the list of subsets
           $Q_{1}^{\sigma_{1}},Q_{2}^{\sigma_{2}},\dots,Q_{d}^{\sigma_{d}}$
   
    \STATE \textbf{Stage 3: Server Processing}
    \STATE\hspace{0.6em}Upon receiving
           $Q_{1}^{\sigma_{1}},\dots,Q_{d}^{\sigma_{d}}$,
           server $\sigma_{1}\!\cdots\!\sigma_{d}$ computes
           \[
             a_{\alpha}=
               \bigoplus_{j_{1}\in Q_{1}^{\sigma_{1}}}\;
               \bigoplus_{j_{2}\in Q_{2}^{\sigma_{2}}}\;
               \cdots\!
               \bigoplus_{j_{d}\in Q_{d}^{\sigma_{d}}}
                 x_{\,j_{1},\dots,j_{d}}
           \]
           and returns the single bit $a_{\alpha}$ to the client
   
    \STATE \textbf{Stage 4: Reconstruction}
    \STATE\hspace{0.6em}(Client) XOR the $k=2^{d}$ replies:
           $\displaystyle
             x_{i_{1},\dots,i_{d}}
               =\bigoplus_{\alpha\in\{0,1\}^{d}} a_{\alpha}$
  \end{algorithmic}
\end{algorithm*}

\textbf{Correctness:}   The scheme's correctness consists of two parts. One is the fact that every
bit in $x$, except $ x_{i}$  appears in an even number of subcubes $ x_{\sigma}, \sigma \in \{0,1\}^{d}$ (and $ x_{i} $ appears in exactly one such subcube).  It is not hard to see that $\left(i_1, \ldots, i_d\right)$ is the only position contained in an odd number of subcubes. Actually position $\left(i_1, \ldots, i_d\right)$ appears in a single subcube. Since for every $t \in[d]$, the value $i_t$ appears in exactly one of the sets $Q_t^0, Q_t^1$. Each of the other positions $\left(j_1, \ldots, j_d\right)$ (i.e., those $\neq\left(i_1, \ldots, i_d\right)$) appears in an even number of subcubes. Therefore,  the contribution of these positions is canceled and the only value that remains is that of position $\left(i_1, \ldots, i_d\right)$.  Another comes from the property of $\text{QFT}^{-1}$ that

$$
\text{QFT}^{-1}\left(\frac{1}{\sqrt{N}} \cdot \sum_{j=0}^{N-1}  e^{2 \pi i  \frac{\sum_{k=1}^n x_k}{N}\cdot j}\right) = |  \sum_{k=1}^n x_k \ \text{mod}\  N \rangle _{c}.
$$

\textbf{Privacy:}
  \begin{enumerate}
   \item \textbf{ Uniformity of single-server query distribution:} For any server, the queries it receives, denoted as $Q$ or $Q'$, are uniformly distributed over the range of the target index $i$. When the user index is $i $, the query pair $(Q,Q')$ adheres to the condition $Q'=Q\oplus i$, with $Q$ being independently and uniformly selected at random. For any server, such as Server 1, which exclusively receives $Q$, the potential values of the target index $i=Q\oplus Q'$ span the entire database index space. Therefore, a single server cannot infer any information about $i$ solely from observing  $Q $.

 \item  \textbf{Infeasibility of multi-server conspiracy:} Up to $t$ servers jointly analyze their query sets and still cannot obtain statistical information about $i$.

   The key  $i=(i_1,\dots,i_d)$ is split into $d$ components, and a corresponding $d$ random set $Q_1^{\sigma_1}, \dots, Q_d^{\sigma_d}$ is sent to each of the $k=2^d$ servers, where $\sigma_j \in \{0,1\}$. Each pair $(Q_j^0, Q_j^1)$ satisfies $Q_j^1 = Q_j^0 \oplus i_j$.

Uniform coverage: for any $t$ conspiratorial servers (corresponding to known combinations of some of the components $\sigma_j$), the query set corresponding to the remaining $d-t$ components remains uniformly random and independent of $\{i_{t+1},\dots,i_d\}$, and the conspirators are required to guess the dissimilarity result of the unknown components with a probability of success of no more than $2^{-(d-t)}$.

 \item \textbf{Non-clonability of Quantum State Privacy:}. The quantum bit string $| j \rangle_{t}$ transmitted by Server 1 is computationally based and does not incorporate the phase $e^{2\pi i \frac{x_Q}{N}j}$. Consequently, Server 2 is unable to utilize this state independently to deduce $x_Q$. In the final global state $| \psi_3\rangle$, the phase parameter $x_Q \oplus x_{Q'}$ is only applicable for Client decoding. The server does not possess the complete superposition state and is therefore unable to ascertain the combined value through local measurements.

\end{enumerate}

\section{Conclusion}
In this paper, we have established a new, tighter lower bound on QPIR communication complexity by formulating privacy and correctness in terms of quantum relative entropy and mutual information, and leveraging Uhlmann’s lemma alongside a quantum Pinsker–type inequality. This bound refines previous classical entropy analyses and quantifies how much communication is fundamentally unavoidable even when allowing for specious (i.e., potentially malicious but indistinguishable) server behavior.

By building on our new tight communication bound, all three QPIR schemes consistently achieve sublinear communication complexity while maintaining information‐theoretic security against specious quantum adversaries. This common advantage—operating below the linear threshold imposed by previous analyses—demonstrates the practical impact of our theoretical bound, as each construction remains compatible with existing QKD infrastructure and is supported by rigorous proofs of privacy and correctness.

Looking ahead, we aim to reduce client‐side overhead—especially in the QHE protocol—by integrating more compact quantum components. We also plan to extend the authentication framework to higher noise levels and partial trust models for NISQ hardware \cite{Preskill18}, and to generalize our multi‐server design to support larger collusion thresholds while preserving sublinear communication. Finally, implementing these schemes on prototype quantum or hybrid platforms will be crucial to validate their practicality and guide further optimizations.

\end{document}